\documentstyle[preprint,prl,aps]{revtex}
\begin{document}
\draft
\title{Unusual low-temperature thermopower
in the\\ one-dimensional Hubbard model}
\author{C.~A.~Stafford}
\address{Center for Superconductivity Research, Department of Physics\\
University of Maryland, College Park, Maryland 20742}
\date{Received 9 March 1993}
\maketitle
\begin{abstract}
The low-temperature thermoelectric power of the repulsive-interaction
one-dimensional Hubbard model is calculated
using an asymptotic Bethe ansatz for holons and spinons.
The competition between the entropy
carried by the holons and that carried by the backflow of the spinons gives
rise to an unusual temperature and doping dependence of the thermopower
which is qualitatively
similar to that observed in the normal state of
high-$T_{c}$ superconductors.
\end{abstract}

\narrowtext
Interacting one-dimensional (1D) electron systems generically exhibit {\em
spin-charge separation} \cite{schulz},
that is, the Hamiltonian separates at low
energies into independent terms describing the charge and spin degrees
of freedom.
To date, the thermopower of interacting 1D electron
systems has only been calculated for a few special cases, where the
spin degrees of freedom reduce to noninteracting spins
\cite{kwakbeni,chaikin},
or where the contribution of the spin excitations to the thermopower is
negligible \cite{schulz}.
A more general result for the thermopower of such systems is of
fundamental interest since the entropy carried by 
the spin excitations
represents a qualitatively new type of thermopower, distinct from
the familiar contributions of charge carrier diffusion, phonon drag, etc.
Furthermore, Anderson has argued that the physics of the
$\mbox{Cu} \mbox{O}_2$ planes in high-$T_{c}$ superconductors
is that of spin-charge separation \cite{phil2d}, and it is an interesting
question whether their unusual
normal state thermopower \cite{exphole,expelectron1,expelectron2,magtherm}
can be explained on that basis.  Since a rigorous treatment of
spin-charge separation in 2D systems is still lacking, it is clearly of
interest to investigate the effects of spin-charge separation on the
thermopower of 1D systems, for which
powerful methods such as the Bethe ansatz are available.

In this letter, we calculate explicitly
the contributions of both the charge and spin excitations to the
low-temperature thermopower of the repulsive-interaction 1D Hubbard model
in two limiting cases:  in the strong-coupling limit, and
near the Mott-Hubbard metal-insulator transition occuring at half filling.
In both of these limits, the charge degrees of freedom of the model
can be mapped onto weakly interacting
spinless fermions \cite{schulz,kwakbeni,ogatashiba,andyandi},
while the spin excitations can be shown to form an {\em ideal semion gas}
\cite{haldanespinongas} at low temperatures, which interacts with the charge
degrees of freedom via a backflow
condition that ensures that the electric current
which flows in response to an electric field
is really a current of {\em electrons}, which carry both charge and spin.
The competition between
the entropy carried by the charge excitations and that carried
by the backflow of spin excitations leads to a
nontrivial temperature and doping dependence of the thermopower.
We comment briefly on the possible relevance of
these results to an understanding of the normal state thermopower of
high-$T_c$ superconductors and quasi-one-dimensional organic conductors.

The thermoelectric power $S$ is given in the Kubo formalism by
\begin{equation}
S = \frac{1}{T} \left(\frac{\int_{0}^{\infty} \langle \hat{J}_E(0)
\hat{J}_e(t) \rangle \,dt}{\int_{0}^{\infty} \langle \hat{J}_e(0)
\hat{J}_e(t) \rangle \, dt} + \frac{\mu}{e}\right),
\label{seebeck}
\end{equation}
where $\hat{J}_e$ and $\hat{J}_E$ are the electric current and energy
current operators in the Heisenberg representation,
$\mu$ is the chemical potential, $e$ is the absolute
value of the electron charge, and $\langle \cdots \rangle$ denotes the
thermal average.  In a system with spin-charge separation, the energy
current can be decomposed into a term associated with charge
excitations and a term associated with spin excitations, $\hat{J}_E =
\hat{J}_{E}^{c} + \hat{J}_{E}^{s}$, and
the chemical potential can be written as $\mu = \mu_c + \mu_s$, where
$\mu_c = \partial E_0 / \partial N + \partial F_c /\partial N$ and
$\mu_s = \partial F_s / \partial N$, $E_0$ being the ground state
energy, $N$ the number of electrons in the system, and $F_c$ and $F_s$ the
free energies of the charge and spin excitations.
The thermopower can therefore be expressed as $S = S_c + S_s$, where
$S_c$ and $S_s$ are defined by Eq.~(\ref{seebeck}) with
$\hat{J}_{E}^{c,s}$ and $\mu_{c,s}$ in place of $\hat{J}_{E}$ and $\mu$.
$S_c$ and $S_s$ can be interpreted as the entropies transported separately
by charge and spin excitations when an electric current flows.

We specialize our arguments to the Hubbard model of
spin-1/2 fermions hopping with matrix element $t$ between
nearest-neighbor sites of a 1D lattice
with unit lattice constant, and subject to a repulsive interaction $U$
when two fermions (of opposite spin) occupy the same lattice site.
Periodic boundary conditions are imposed on the lattice, which
consists of $L$ sites, and the system is threaded by
a time-dependent magnetic flux $(\hbar c/e)\Phi(t)$.  (In the following,
we set $\hbar = 1$.)  The complete excitation spectrum of this model
was obtained by Woynarovich \cite{woy4} for the case $\Phi(t) = 0$;
we extend the results of Ref.~\cite{woy4} to $\Phi(t) \neq 0$ using the
arguments of Ref.~\cite{ss}.  In the large-$L$ limit, the energy
of the system with $N=L-H$ electrons can be expressed as \cite{woy4}
\begin{equation}
E(L-H)=E_0(L) -\sum_{h=1}^{H} \varepsilon_{c}(k_h) +
\sum_{\sigma =1}^{N_{s}} \varepsilon_{s}(\Lambda_{\sigma}),
\label{enerholes}
\end{equation}
where $E_0(L)$ is the ground state energy at $N=L$, obtained
in Ref.~\cite{lwu}, $\varepsilon_{c}(k) = -2 t \cos k  -
4 t \int_{0}^{\infty} \! d\omega J_1(\omega) \cos (\omega \sin k) /
[\omega  + \omega \exp(\omega U/2t)]$, and
$\varepsilon_{s}(\Lambda) =
2 t \int_{0}^{\infty} \! d\omega
J_1(\omega) \cos(\omega \Lambda)\, \mbox{sech}(\omega U/4t) /\omega$.
The momentum [defined modulo($2\pi$)] is given by \cite{woy4}
\begin{equation}
P = \Phi(t) - \sum_{h=1}^{H} p_{c}(k_h) +
\sum_{\sigma =1}^{N_{s}} p_{s}(\Lambda_{\sigma}) + \psi,
\label{pholes}
\end{equation}
where $\psi = \pi (L- N/2 + N_s/2 +1)$,
$p_{c}(k) = k + 2 \int_{0}^{\infty} \! d\omega  J_{0}(\omega)
\sin (\omega \sin k) / [\omega + \omega \exp(\omega U/2t)]$, and
$p_{s}(\Lambda) = - \int_{0}^{\infty} \! d\omega J_0 (\omega)
\sin (\omega \Lambda)\, \mbox{sech}(\omega U/4t) /\omega$.
The parameters $k_h$ and $\Lambda_{\sigma}$
are real numbers satisfying $-\pi \leq k_h \leq \pi$,
$-\infty \leq \Lambda_{\sigma} \leq \infty$, and can be interpreted as
holes in the ground state distributions of {\em pseudomomenta} and {\em
spin rapidities} in the Lieb-Wu equations \cite{lwu}; such excitations
are referred to as {\em holons} and {\em spinons}.
(In Eq.~(\ref{enerholes}), we have
omitted states with complex pseudomomenta, for which there is a finite
energy gap \cite{woy4}.)
The holons and spinons can not in general be regarded as noninteracting
quasi-particles since the $k_h$ and $\Lambda_{\sigma}$ are not free
parameters, but are determined by
the asymptotic Bethe ansatz equations derived in Ref.~\cite{woy4}:
\widetext
\FL
\begin{eqnarray}
L p_{c}(k_h) & = & 2\pi I_h + \Phi(t) +
\sum_{h'=1}^{H} \Theta_1 (\sin k_h -\sin k_{h'})
+\sum_{\sigma =1}^{N_{s}} \Theta_2 (\sin k_h - \Lambda_{\sigma})
- \pi \sum_{\alpha =1}^{M_{s}} \mbox{sgn} (\sin k_h - \mbox{Re}
\lambda_{\alpha}),
\label{holonansatz}\\
L p_{s}(\Lambda_{\sigma}) & = & -2\pi J_{\sigma} - \sum_{h=1}^{H}
\Theta_2 (\Lambda_{\sigma}-\sin k_h) + \sum_{\sigma'=1}^{N_{s}}
\Theta_1(\Lambda_{\sigma}-\Lambda_{\sigma'}) - \sum_{\alpha=1}^{M_{\rm
s}} 2 \tan^{-1} \frac{\Lambda_{\sigma}-\lambda_{\alpha}}{U/4t},
\label{spinonansatz}
\end{eqnarray}
\narrowtext
\vspace{-5mm}
\FL
\begin{equation}
\sum_{\sigma =1}^{N_{s}} 2
\tan^{-1}\frac{\lambda_{\alpha}-\Lambda_{\sigma}}{U/4t} = 2\pi K_{\alpha}
+ \sum_{\beta =1}^{M_{s}} 2\tan^{-1} \frac{\lambda_{\alpha} -
\lambda_{\beta}}{U/2t},
\label{secondspinonansatz}
\end{equation}
where $M_{s}$ is the number of down-spin spinons, and the two-body
scattering phase shifts are given by
$\Theta_1 (x)=
2 \int_{0}^{\infty} d\omega \sin(\omega x)
/ [\omega + \omega \exp(\omega U/2t)]$,
$\Theta_2(x) = 2 \tan^{-1} [\tanh(\pi t x/U)]$.
The charge degrees of freedom are specified by the $H$ quantum numbers
$I_h$ [defined modulo($L$)], which
are distinct integers (half-integers) for $M$ even (odd), $M$ being
the number of down-spin electrons.  The spin degrees of freedom are
specified by the $N_s$ distinct quantum numbers $J_{\sigma}$, which take
values in the range $-J_{\rm max}$, $-J_{\rm max} + 1$, $\ldots$,
$J_{\rm max}$, where $J_{\rm max} = (N-M-1)/2$,
and by the $M_s$ quantum numbers
$K_{\alpha}$, which are integers (half-integers) for $N_{s}-M_{s}$ odd (even).
The $I_h$ and $J_{\sigma}$ represent holes in the ground state
distributions of charge and spin quantum numbers in the Lieb-Wu equations,
while the $K_{\alpha}$ and the complex parameters
$\lambda_{\alpha}$ describe the string-like states in
the spin sector of the model (their relation to the parameters of
the Lieb-Wu equations is given in Ref.~\cite{woy4}).
The $z$-component of the total spin of the system is $S^z = (N-2M)/2 =
(N_{s} -2 M_{s})/2$.  For the ground state in zero magnetic field,
the $I_h$ are
consecutive integers (or half-integers) centered about $L/2$ and $N_{s} = N
\bmod 2$. 

To evaluate the thermopower (\ref{seebeck}) using the above
formalism would be quite difficult in general because, while the
chemical potential can be extracted from the energy spectrum
(\ref{enerholes}), the matrix elements of the current
operators would have to be evaluated using the Bethe ansatz
wavefunctions \cite{lwu,ogatashiba}, which are quite unwieldy.
We therefore
consider two limiting cases where the weakness of the holon-holon and
spinon-spinon
interactions allows one to construct the matrix elements of the current
operators explicitly.  Both for $H\ll L$ and for $U\gg t$, the holons
can be mapped onto weakly interacting spinless fermions
\cite{schulz,kwakbeni,ogatashiba,andyandi},
allowing the matrix elements of $\hat{J}_e$ and
$\hat{J}_{E}^{c}$ to be evaluated; in addition, for both of these cases
one can show that in the low-temperature limit ($k_B T \ll v_s$) the
energy spectrum of the spin excitations is
\begin{equation}
E_s \simeq v_s \sum_{\sigma =1}^{N_s} \left( \frac{\pi (1-\delta)}{2} -
\left|\frac{2 \pi J_{\sigma}'}{L}\right|\right)
+ \frac{\pi v_s N_{s}^2}{4 L},
\label{idealspinons}
\end{equation}
where $v_s$ is the spinon velocity in the low-energy limit,
calculated in Refs.~\cite{coll,schulz}, $\delta \equiv |1-N/L|$,
and $J_{\sigma}' = J_{\sigma} + \mbox{sgn}(J_{\sigma}) M_s /2$.
Eq.~(\ref{idealspinons}) follows upon linearizing the spinon dispersion
relation in the low-energy ($\Lambda \rightarrow \pm \infty$) limit,
replacing the spinon-spinon scattering phase shifts by their limiting
low-energy form
$\Theta_1 (\Lambda_{\sigma} -\Lambda_{\sigma'}) \simeq
(\pi/2)\mbox{sgn}(\Lambda_{\sigma} - \Lambda_{\sigma'})$, and taking
$\tan^{-1}[4t(\Lambda_{\sigma} -\lambda_{\alpha})/U] \simeq
(\pi/2)\mbox{sgn}(\Lambda_{\sigma})$.  The low-energy spectrum
(\ref{idealspinons}) is
equivalent to that of the ``ideal spinon gas'' described in
Ref.~\cite{haldanespinongas}:
for fixed $N_s$, the degeneracies are equivalent to
those of a system of spin-1/2 bosons, while the behavior of the
Hilbert space as $N_s$ is varied implies that the spinons are in fact
{\em semions}, the second term in Eq.~(\ref{idealspinons}) representing
the statistical interaction.
The special feature of Haldane's ideal spinon gas is that the
spinon-spinon interactions are described exactly by mean-field theory,
so that spin exchange processes between spinons are absent
\cite{haldanespinongas}.  Such processes are implicit in
Eqs.~(\ref{spinonansatz}) and (\ref{secondspinonansatz}), which can be
thought of as a nested Bethe-Yang ansatz for the spinons.
However, spinon spin exchange processes have vanishing amplitude
in the low-energy limit,
so that the low-energy spectrum of the spin excitations (\ref{idealspinons})
is independent of the quantum numbers $K_{\alpha}$ which specify
the spin wavefunction of the spinons.
The excitation spectrum (\ref{idealspinons}) implies
a low-temperature spin entropy per site of $\pi k_{B}^2 T / 3 v_s$
\cite{haldanespinongas}, from which it follows that
\begin{equation}
\lim_{T \rightarrow 0} \mu_s = \frac{\pi (k_B T)^2}{6
v_{s}^2} \frac{\partial v_s}{\partial n},
\label{muspin}
\end{equation}
where $n \equiv N/L$.  $\mu_s$ is not to be confused with the spinon
chemical potential, which is zero since spinons are thermal excitations.

Because the spinon-spinon interactions have the mean-field form
(\ref{idealspinons}) in the low-energy limit, they do not
contribute to the spinon energy current $\hat{J}_{E}^s$.
Consequently, $\hat{J}_{E}^s$ commutes with
the Hamiltonian in the low-energy limit, and has the eigenvalue
$J_{E}^s = v_{s}^2 P_s$ (assuming equal numbers of right and left
movers), where $P_s$ is the total momentum of the spinon gas.
To calculate the contribution of the spinon energy current to the
thermopower,  we make use of the Onsager relation between the
thermopower and the Peltier coefficient, and define $\Pi^s \equiv T S_s
= \Pi_{E}^s + \mu_s/e$, where $\Pi_{E}^s =
\langle \hat{J}_{E}^s \rangle / \langle \hat{J}_e \rangle$ evaluated for
a system subject to an infinitesimal electric field
${\cal E} = -L^{-1} d\Phi /d t$ and with $\nabla T = 0$.
The electric field is equivalent to a time-dependent shift of the holon
quantum numbers $I_h$ in Eq.~(\ref{holonansatz}), which leads to a shift
in the $k_h$ distribution and an electric current
$J_e = \langle \hat{J}_e \rangle$ proportional to ${\cal E}$.  (The
system has a finite dc conductivity at $T > 0$ due to activated Umklapp
processes which degrade the electric current; however, the ratio of
current-current correlation functions in Eq.~(\ref{seebeck}) is
independent of the Umklapp scattering rate in the low-temperature limit.)
The shift in the $k_h$ distribution leads via the spinon-holon
scattering phase shifts $-\Theta_2 (\Lambda_{\sigma} -\sin k_h)$ in
Eq.~(\ref{spinonansatz}) to a shift in the $\Lambda_{\sigma}$
distribution.  This backflow of spinons results because the
electric field really couples to electrons, which carry both charge and spin.
Both for $H \ll L$ and for $U \gg t$, one can approximate
$\Theta_2 (\Lambda_{\sigma} - \sin k_h) \simeq \Theta_2
(\Lambda_{\sigma}) - \mbox{sech}(2\pi t \Lambda_{\sigma}/U)\,(2 \pi t\sin
k_h/U)$; $\Theta_2 (\Lambda_{\sigma})$ is absorbed in a redefinition of
the spinon momentum at finite $H$:  $\tilde{p}_s (\Lambda_{\sigma}) = p_s
(\Lambda_{\sigma}) + H \Theta_2 (\Lambda_{\sigma})/L$.  Summing
over $\tilde{p}_s (\Lambda_{\sigma})$, we obtain
\begin{equation}
\langle \hat{J}_{E}^s \rangle
= v_{s}^2 \left\langle \frac{1}{L} \sum_{\sigma
=1}^{N_s} \frac{2 \pi t}{U} \mbox{sech}\frac{2 \pi t
\Lambda_{\sigma}}{U} \right\rangle_{\! J_e = 0}
\left\langle \sum_{h=1}^{H} \sin k_h \right\rangle_{\! J_e}\!,
\label{spinonenergycurrent}
\end{equation}
which we evaluate explicitly below, using the fact that the sum over $k_h$ is
proportional to $J_e$, while the sum over $\Lambda_{\sigma}$
is proportional to the excitation energy of the spinon gas,
which can be evaluated via the correspondence with the
model of Ref.~\cite{haldanespinongas}.

We first consider the limit $\delta \equiv |1-n| \ll 1$,
close to the Mott-Hubbard metal-insulator transition.  This case has been
studied previously via a weak-coupling approximation \cite{schulz};
we extend the results of Ref.~\cite{schulz} to arbitrary $U$, and
explicitly verify the assertion of Ref.~\cite{schulz} that the contribution
of the spin excitations to the thermopower is negligible in this limit.
Eqs.~(\ref{enerholes}) and (\ref{pholes}) implicitly define an energy
band $\varepsilon_c (k(p))$ for charge excitations, $k(p)$ being the
inverse of the function $p_c (k)$.  The holons can be thought of as
holes in this energy band, which may be approximated
near the energy minimum at $p=\pi$ by
$\varepsilon_{c}(k(p)) \simeq \mu_- - (p-\pi)^2/2|m^{\ast}|$,
where $\mu_- =\varepsilon_{c}(\pi)$ is the $T=0$
chemical potential in the limit $n \rightarrow 1^-$ \cite{lwu}, and
\[
|m^{\ast}| = \frac{1}{2t}
\frac{\{1-2\int_{0}^{\infty} d\omega J_{0}(\omega)/
[1 + \exp(\omega U/2t)] \}^2}{
1-2\int_{0}^{\infty} d\omega \, \omega J_{1}(\omega)/
[1 + \exp(\omega U/2t)]}
\]
is the absolute value of the holon effective mass \cite{kawokiji}.
Eq.~(\ref{holonansatz}) implies that the holon momenta $p_{c}(k_h)$
differ from those of noninteracting spinless fermions by a term which
vanishes as $H/L$, $N_{s}/L \rightarrow 0$;
we write $p_h \equiv p_{c}(k_h) = 2\pi I_{h}/L + \Phi/L + \delta p_{h}$,
where $\delta p_h$ has contributions from holon-holon, holon-spinon, and
holon-string scattering.  Holon-string scattering merely shifts the
parity of the holon quantum numbers $I_h$, and is therefore unimportant
in the limit $L \rightarrow \infty$, while holon-holon and holon-spinon
scattering give $\delta p_h/(p_h - \pi) \rightarrow
-4 \ln 2 \, \delta/U p_{c}'(\pi) +
C T^2$ as $\delta$, $T \rightarrow 0$, where $C$ is a $U$-dependent
constant.  The shifts of the holon momenta due to the interactions thus
have a negligible effect on the charge excitation energies in the limit
$\delta$, $T \rightarrow 0$.
Because of the vanishing interactions, the current operators $\hat{J}_e$
and $\hat{J}_{E}^c$ commute with the Hamiltonian in the limit $\delta$,
$T \rightarrow 0$, and Eq.~(\ref{seebeck}) can be evaluated
straightforwardly to give
\begin{equation}
\lim_{\delta, T \rightarrow 0} S_c
= \mbox{sgn}(1-n) \, \frac{k_{B}^{2} T}{3e}
\frac{|m^{\ast}|}{\delta^2},
\label{Snearhalfilling}
\end{equation}
where we have used the electron-hole symmetry of the model about $n=1$
\cite{woy4,andyandi}.  The corrections to Eq.~(\ref{Snearhalfilling})
due to holon-holon interactions are expected to be ${\cal O}(\delta^{-1})$.
Evaluating Eqs.~(\ref{muspin}) and (\ref{spinonenergycurrent}) to obtain
$\mu_s$ and $\Pi_{E}^s$, we obtain
\begin{equation}
\lim_{\delta, T \rightarrow 0} S_s  =
\mbox{sgn}(n-1) \frac{\pi k_{B}^2 T}{12 e t} \left[ \frac{1}{I_1 (2 \pi
t/U)} - \frac{4 \pi t^2 |m^{\ast}|}{U \, p_{c}'(\pi) I_0 (2 \pi t/U)} \right],
\end{equation}
where $I_0$ and $I_1$ are modified Bessel functions.
$S_s$ is negligible compared to $S_c$ in the limit $\delta \rightarrow 0$.
The low-temperature thermopower of the system thus becomes
large and positive (hole-like) as the metal-insulator transition is
approached from $n<1$, and has the opposite sign (electron-like) as $n
\rightarrow 1$ from above.

In the limit $U \gg t$, the
full $n$-dependence of the low-temperature thermopower can be
calculated:  When $U / t \rightarrow \infty$,
$\varepsilon_c (k) \rightarrow -2 t \cos k$, $p_c (k) \rightarrow k$, and the
scattering phase shifts in Eq.~(\ref{holonansatz}) are ${\cal O}(t/U)$,
leading to the well-known mapping of the holons onto noninteracting
spinless fermions in the strong-coupling limit
\cite{kwakbeni,ogatashiba}.  The spinon dispersion relation
now explicitly involves a contribution from the backflow of the holon
distribution; Eqs.~(\ref{enerholes})--(\ref{secondspinonansatz}) give
$v_s = (2\pi t^2/U)
[1 - \sin (2 \pi n)/2 \pi n - 8 \ln 2 (t/\pi U)
\sin^3 \pi n] + {\cal O}(t^4/U^3)$,
which is consistent with the results of Refs.~\cite{coll,schulz}.
The dominant contributions to the low-temperature
thermopower come from $S_c$ and $\mu_s /e T$, which
combine to give (for $n<1$)
\widetext
\FL
\begin{equation}
\lim_{\stackrel{\scriptstyle U \rightarrow \infty}{T \rightarrow 0}}
S = - \frac{k_{B}^2 T}{e t}
\left(\frac{\pi^2 \cos \pi n}{6 \sin^2 \pi n}\right)
- \frac{k_{B}^2 T}{3 e J} \left(\frac{\cos (2 \pi n)/n -
\sin (2 \pi n) /2 \pi n^2
+ 24 \ln 2 \, (t/U) \sin^2 \pi n \cos \pi n}{
[1 - \sin(2 \pi n)/2 \pi n - 8 \ln 2 \, (t/\pi U) \sin^3 \pi n]^2}\right),
\label{SforlargeU}
\end{equation}
\narrowtext
\noindent
where $J= 4t^2/U$ is the antiferromagnetic superexchange constant and
$S(2-n) = - S(n)$.  The corrections to Eq.~(\ref{SforlargeU}) are
${\cal O}(k_{B}^2 T/e U)$ as $T \rightarrow 0$, and come from
holon-holon interactions, higher order terms in $v_s$, and from
Eq.~(\ref{spinonenergycurrent}), which gives $\Pi_{E}^s/T =
\pi^2 k_{B}^2 T/6 e U n$.
In the physically interesting parameter range $100 > U/t \gg 1$,
Eq.~(\ref{SforlargeU}) implies that the
low-temperature thermopower is positive for all $n < 1$ and negative
for all $n > 1$.
The full temperature dependence of $S_c$ [first term in
Eq.~(\ref{SforlargeU})]
can be calculated by the method of Ref.~\cite{kwakbeni} (neglecting terms
of order $k_B T/U$); as $T$ increases, $S_c$
monotonically approaches the high-temperature value
$\mbox{sgn}(1-n)(k_B/e) \ln[(1-\delta)/\delta]$ \cite{chaikin}.
The full temperature dependence of $S_s$
[second term in Eq.~(\ref{SforlargeU})] is an open problem; however,
when $k_B T \gg J$, $\mu_s$ is dominated by the spin entropy and
$e \, \Pi_{E}^s / k_B T \ll 1$, so that $S_s \rightarrow \mbox{sgn}(n-1)
(k_B/e) \ln 2$, in agreement with the result of Ref.~\cite{kwakbeni}.
Combining the high-temperature results for $S_c$ and $S_s$ yields the
well-known Heikes formula \cite{chaikin}.
For small hole dopings $\delta \ll 1$, $S$ is
dominated by $S_c$, which is large, positive, and a monotonic function
of temperature.  However, for $\delta > 1/3$, $S$ is
dominated by $S_s$, and is negative at high temperatures, but with a
positive slope at low temperatures, implying the existence of a
{\em positive peak} in the low-temperature thermopower.
$S$ has the opposite sign for electron doping.

The magnetic field dependence of $S$ is also readily obtained in the
strong-coupling limit:  a weak field $B$ applied
parallel to the chain does not couple to the charge degrees of freedom
in the Peierls approximation, but
the Zeeman coupling leads for $\mu B \ll k_B T \ll J$ to
\begin{equation}
\Delta S(B) = \frac{k_B}{e} \frac{k_B T}{\pi v_s} \,
\frac{\partial \ln v_s}{\partial n} \left(\frac{\mu B}{k_B T}\right)^2.
\label{magnetothermopower}
\end{equation}

It is interesting to compare the temperature and doping
dependence of the thermopower in the 1D Hubbard model with that observed
in the cuprate materials in which high-$T_c$ superconductivity occurs,
which are widely regarded to be quasi-two-dimensional
doped Mott insulators \cite{phil2d}.
The in-plane thermopower of the lightly doped cuprates is generically positive
for hole doping \cite{exphole} and negative for electron doping
\cite{expelectron1,expelectron2},
with a magnitude which increases drastically as the nominal concentration
of doped carriers goes to zero, in qualitative agreement with
Eq.~(\ref{Snearhalfilling}).
Upon further hole doping, the in-plane thermopower of the cuprates universally
exhibits a positive peak at low temperatures, then
decreases monotonically, often becoming negative at room temperature in
the superconducting samples \cite{exphole};
the mirror image behavior, with a negative peak
at low temperatures, is exhibited \cite{expelectron2}
by the electron-doped superconductor $\mbox{Nd}_2
\mbox{Cu}\mbox{O}_{4-x}\mbox{F}_x$.  Similarly, the unusual temperature
dependence of the spinon backflow thermopower in
the 1D Hubbard model leads for $\delta > 1/3$ and $U\gg t$
to a low-temperature peak in the thermopower
which is positive for hole doping and negative for electron doping.
Superconducting samples \cite{expelectron1} of
$\mbox{Nd}_{2-x} \mbox{Ce}_x \mbox{Cu}\mbox{O}_4$ exhibit a peak of the same
sign as that observed in the hole-doped cuprates, however, which can not be
accomodated in a model with electron-hole symmetry, such as the Hubbard model.
It is of interest, however, that the normal-state thermopower of
$\mbox{Nd}_{2-x} \mbox{Ce}_x \mbox{Cu}\mbox{O}_4$
goes to zero as $T\rightarrow
0$ when a magnetic field is applied to suppress the superconductivity
\cite{magtherm}, which is consistent with the linear low-temperature
thermopower we calculate.  The smallness of
the isotropic contribution to the magnetothermopower \cite{magtherm}
is qualitatively consistent with our result for the 1D Hubbard
model, Eq.~(\ref{magnetothermopower}), which is reduced in
magnitude by a factor of order $k_B T/J$ compared to the high-temperature
value \cite{chaikin2}.
The spin contribution to the thermopower, which takes the value
$-(k_B/e)\ln 2$ in the high-temperature limit,
has been previously identified \cite {chaikin2} in
a family of quasi-one-dimensional organic conductors
which have been modeled
as Hubbard chains with $n=1/2$ and $U \gg t$;
from Eq.~(\ref{SforlargeU}), we would expect the low-temperature
thermopower of these systems to be positive, and
a recent measurement \cite{Jerome} on a closely related compound indeed
exhibits a positive peak at low temperatures.

It is a pleasure to thank P.~W.~Anderson, S.~Das Sarma, J.~H.~Kim,
and A.~J.~Millis for valuable discussions.
This work was supported by NSF grant DMR-91-23577.

\tighten


\begin{references}

\bibitem{schulz} H.~J.~Schulz, Int. J. Mod. Phys. B {\bf 5}, 57 (1991).

\bibitem{kwakbeni} J.~F.~Kwak and G.~Beni, Phys. Rev. B {\bf 13}, 652 (1976).

\bibitem{chaikin} P. M. Chaikin and G. Beni, Phys. Rev. B {\bf 13}, 647
(1976).

\bibitem{phil2d} P.~W.~Anderson, Phys. Rev. Lett. {\bf 64}, 1839 (1990);
{\bf 65}, 2306 (1990); {\bf 66}, 3226 (1991); {\bf 67}, 2092 (1991).

\bibitem{exphole} A. B. Kaiser and C. Uher, in {\em Studies of High
Temperature Superconductors}, Vol. 7, edited by A. V. Narlikar (Nova Science
Publishers, New York, 1990); S. D. Obertelli, J. R. Cooper, and J. L.
Tallon, Phys. Rev. B {\bf 46}, 14928 (1992).

\bibitem{expelectron1} X.~Q.~Xu {\em et al.}, Phys. Rev. B {\bf 45}, 7356
(1992).

\bibitem{expelectron2}
J.~Sugiyama {\em et al.}, Phys. Rev. B {\bf 45}, 9951 (1992).

\bibitem{magtherm}
W.~Jiang, X.~Q.~Xu, S. J. Hagen, J. L. Peng, Z. Y. Li, and R. L.
Greene (to be published).

\bibitem{ogatashiba} M.~Ogata and H.~Shiba, Phys. Rev. B {\bf 41}, 2326
(1990).

\bibitem{andyandi} C.~A.~Stafford and A.~J.~Millis (to be published).

\bibitem{haldanespinongas} F.~D.~M.~Haldane, Phys. Rev. Lett. {\bf 66},
1529 (1991).

\bibitem{woy4} F. Woynarovich, J. Phys. C {\bf 15}, 85 (1982);
{\bf 15}, 97 (1982); {\bf 16}, 5293 (1983).

\bibitem{ss} B.~S.~Shastry and B.~Sutherland, Phys. Rev. Lett. {\bf 65},
243 (1990).

\bibitem{lwu} E. H. Lieb and F. Y. Wu, Phys. Rev. Lett. {\bf 20}, 1445 (1968).

\bibitem{coll}
C. F. Coll, Phys. Rev. B {\bf 9}, 2150 (1974).

\bibitem{kawokiji}
N. Kawakami and A. Okiji, Phys. Rev. B {\bf 40}, 7066 (1989);
J. Carmelo, P. Horsch, P. A. Bares, and A. A.
Ovchinnikov, {\em ibid.} {\bf 44}, 9967 (1991).

\bibitem{chaikin2} P. M. Chaikin, J. F. Kwak, and A. J. Epstein, Phys.
Rev. Lett. {\bf 42}, 1178 (1979).

\bibitem{Jerome} W. Kang, D. Jerome, L. Valade, and P. Cassoux, Synthetic
Metals {\bf 42}, 2343 (1991).

\end{references}
\end{document}